\documentclass[sigconf]{acmart}

\AtBeginDocument{%
  \providecommand\BibTeX{{%
    \normalfont B\kern-0.5em{\scshape i\kern-0.25em b}\kern-0.8em\TeX}}}
    
\usepackage{cleveref}
\usepackage{enumitem}
\usepackage{xspace}
\usepackage{multirow}

\usepackage[ruled,vlined,lined,commentsnumbered]{algorithm2e}
\usepackage{booktabs}
\usepackage{moreverb}
\usepackage{fontenc}
\usepackage{amsmath}

\usepackage{amssymb}
\usepackage{fancybox}
\usepackage{color}
\usepackage{colortbl}
\usepackage{array}
\usepackage{diagbox}
\usepackage{flushend}
\usepackage{multicol}
\usepackage{listings}
\usepackage{makecell}
\usepackage{graphicx}
\usepackage{setspace}
\usepackage{soul}
\usepackage{csvsimple}
\usepackage{longtable}
\usepackage[skip=1pt,font=small,labelfont=bf]{caption}
\usepackage{url}
\usepackage{lscape}
\usepackage{rotating}
\usepackage{tikz}
\usepackage{enumitem}
\usepackage{subcaption}
\usepackage{wrapfig}
\usepackage[most]{tcolorbox}
\makeatletter
\@namedef{ver@lineno.sty}{9999/12/31}
\@namedef{opt@lineno.sty}{}
\makeatother

\copyrightyear{2024} 
\acmYear{2024} 
\setcopyright{acmlicensed}\acmConference[ICSE '24]{2024 IEEE/ACM 46th International Conference on Software Engineering}{April 14--20, 2024}{Lisbon, Portugal}
\acmBooktitle{2024 IEEE/ACM 46th International Conference on Software Engineering (ICSE '24), April 14--20, 2024, Lisbon, Portugal}
\acmDOI{10.1145/3597503.3639120}
\acmISBN{979-8-4007-0217-4/24/04}

\newcommand{\method}{SEC\xspace}

{ \newcommand{\mynote}[2]{
      \fbox{\bfseries\sffamily\scriptsize#1}
        {\small$\blacktriangleright$\textsf{\emph{#2}}$\blacktriangleleft$}}}
        { \newcommand{\mynote}[2]{}}

\begin{document}


\title[When Neural Code Completion Models Size up the Situation]{When Neural Code Completion Models Size up the Situation: Attaining Cheaper and Faster Completion through Dynamic Model Inference}

\author{Zhensu Sun}\authornote{Both authors contributed equally to this research.}
\affiliation{
  \institution{Beihang University}
  \city{Beijing}
  \country{China}
}
\email{zhensuuu@gmail.com}

\author{Xiaoning Du}\authornotemark[1]
\affiliation{
  \institution{Monash University}
  \city{Melbourne}
  \state{Victoria}
  \country{Australia}
}
\email{xiaoning.du@monash.edu}

\author{Fu Song}
\affiliation{%
  \institution{State Key Laboratory of Computer Science, Institute of Software, Chinese Academy of Sciences}
  \city{Beijing}
  \country{China}
}
\additionalaffiliation{
\institution{University of Chinese Academy of Sciences, and Nanjing Institute of Software Technology}
}
\email{songfu@ios.ac.cn}

\author{Shangwen Wang}
\affiliation{
  \institution{National University of Defense
Technology}
  \city{Changsha}
  \country{China}
}
\email{wangshangwen13@nudt.edu.cn}

\author{Li Li}\authornote{Corresponding author.}
\affiliation{%
  \institution{Beihang University, Beijing}
  \city{Yunnan Key Laboratory of Software Engineering}
  \country{China}
}
\email{lilicoding@ieee.org}

\begin{abstract}
Leveraging recent advancements in large language models, modern neural code completion models have demonstrated the capability to generate highly accurate code suggestions.
However, their massive size poses challenges in terms of computational costs and environmental impact, hindering their widespread adoption in practical scenarios.
Dynamic inference emerges as a promising solution, as it allocates minimal computation during inference while maintaining the model's performance.
In this research, we explore dynamic inference within the context of code completion.
Initially, we conducted an empirical investigation on GPT-2, focusing on the inference capabilities of intermediate layers for code completion.
We found that 54.4\% of tokens can be accurately generated using just the first layer, signifying significant computational savings potential.
Moreover, despite using all layers, the model still fails to predict 14.5\% of tokens correctly, and the subsequent completions continued from them are rarely considered helpful, with only a 4.2\% Acceptance Rate.
These findings motivate our exploration of dynamic inference in code completion and inspire us to enhance it with a decision-making mechanism that stops the generation of incorrect code.
We thus propose a novel dynamic inference method specifically tailored for code completion models.
This method aims not only to produce correct predictions with largely reduced computation but also to prevent incorrect predictions proactively.
Our extensive evaluation shows that it can averagely skip 1.7 layers out of 16 layers in the models, leading to an 11.2\% speedup with only a marginal 1.1\% reduction in ROUGE-L.
\end{abstract}

\maketitle

\section{Introduction}

\begin{figure}[t]
\centerline{\includegraphics[width=1\columnwidth]{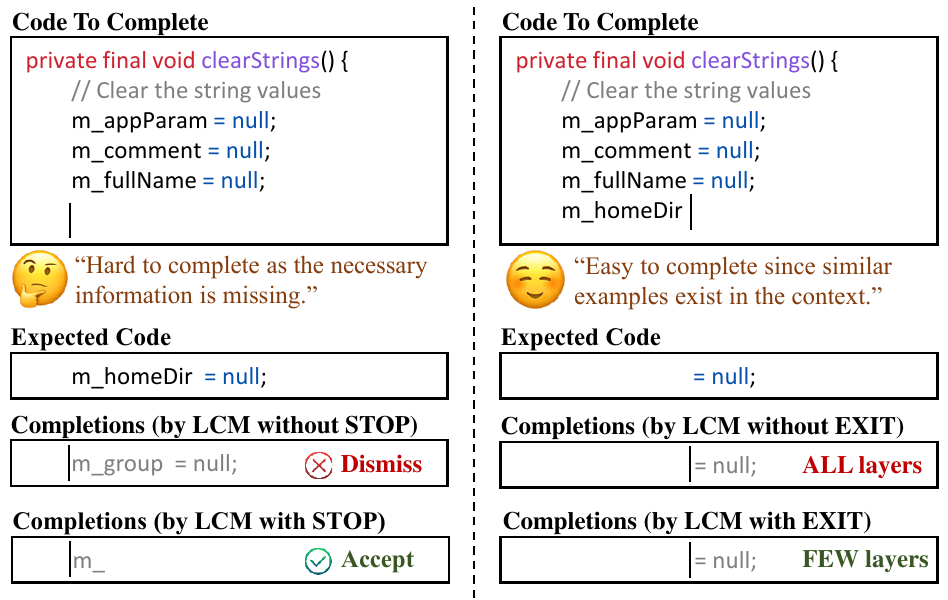}}
\caption{Examples to demonstrate the effects of \method, where the left and right of the figure respectively demonstrate the scenario of STOP and EXIT.}\vspace{-4mm}
\label{fig:example}
\end{figure}

As one of the core functionalities of modern integrated development environments (IDEs), code completion significantly boosts developer productivity by predicting the next few tokens that are likely to be implemented for a given code context.
Recently, pre-trained large language models (LLMs) for code, i.e., Large Code Models (LCMs), have achieved remarkable performance in the code completion task~\cite{hou2023large}.
Motivated by the strong product-market fit and immense business value, a large number of LCM-based commercial code completion applications have been released, including Github Copilot~\cite{copilot}, Cursor~\cite{cursor}, and Amazon CodeWhisperer~\cite{codewhisperer}.

LLMs, with their billions of parameters, necessitate deployment on expensive GPU servers, resulting in substantial financial costs (e.g., approximately \$700,000 per day for ChatGPT~\cite{chatgptCost}) and significant environmental impacts due to high energy consumption and CO$_2$ emission (e.g., 12,800 metric tons of $CO_2$ per year for GPT-3~\cite{chien2023reducing}).
In code completion tasks, the LCM services are usually integrated into IDEs, meaning that the completion requests are automatically issued whenever a typing pause is detected by the IDE, resulting in more frequent and intensive requests compared to typical LLM-based conversational services.
This elevated demand for LCMs in IDEs exacerbates the financial and environmental impacts. 
Moreover, it can limit the practical utilization and commercialization of LCM-based code completion applications.

Dynamic inference~\cite{Schuster2022ConfidentAL,XianggenLiu2020FindingDJ,SuratTeerapittayanon2016BranchyNetFI,XinDai2020EPNetLT,SamLeroux2017TheCN} is one of the most promising
techniques to mitigate this urgent issue associated with large models.
It allows the model to skip some computation and directly output the result when the finished computation suffices to produce a correct result. 
For example, DeeBert~\cite{Xin2020DeeBERTDE} can achieve a 24\% speedup in the inference for the sentiment analysis task without sacrificing accuracy.
However, the performance of the dynamic inference methods shows significant diversity in different downstream tasks.
When applied to textual entailment recognition, DeeBert only achieves a 9\% speedup but comes at the cost of a 0.6\% reduction in accuracy~\cite{Xin2020DeeBERTDE}.
Thus, despite the promising potential advantage of dynamic inference, 
whether and to what extent the code completion systems can benefit from the dynamic inference methods is yet to be explored.

As a starting point, we conducted an empirical study (refer to~\Cref{sec:inv-rq1} for more details) on GPT-2~\cite{Radford2019LanguageMA}, a 12-layer popularly used LCM fine-tuned with Java code snippets. Our investigation aimed to reveal the minimum number of layers required to achieve accurate predictions.
To gain insights into the dynamics of individual layers' contributions to the prediction process, we evaluated GPT-2 on a simplified task of next token prediction instead of generating complete code completions. Surprisingly, we discovered that 54.4\% of tokens can be correctly generated using just the first layer, indicating that a significant portion of the computing resources utilized by the LCM is unnecessary when all layers are always used.
It highlights the importance of dynamic inference in the context of code completion.

Additionally, we observed that GPT-2 fails to correctly predict 14.5\% of tokens even with all its 12 layers.
A manual inspection was conducted to determine the acceptance of continued completions for these wrongly predicted tokens. The results showed a low acceptance rate of 4.2\%, significantly lower than the 46.7\% acceptance rate for completions generated from randomly sampled code contexts.
The low helpfulness of these continued completions for incorrectly predicted tokens implies that generating them wastes computational resources.
Presenting such incorrect predictions could also be misleading for developers and hinder their productivity~\cite{Upadhyaya2022ExpectationVE}.

The encouraging findings from our study inspired us to explore the application of dynamic inference in code completion tasks.
Meanwhile, we also identified that current dynamic inference methods can be further enhanced to better suit the specific characteristics of code completion tasks.
Existing methods are primarily tailored for non-collaborative tasks, such as classification, and consequently, they always produce an output for the user.
However, code completion is a collaborative process involving human intervention, where the requests are automatically issued, and the developer needs to evaluate and integrate the generated suggestions.
Unfortunately, some of the completions generated by these methods may not be helpful and could significantly impede the completion process, leading to negative consequences~\cite{Sun2022LearningTP, Mozannar2023WhenTS}.
From the perspective of developers, having nothing displayed is preferable to receiving unhelpful and time-wasting completions.
To this end, we believe that the dynamic inference method designed for the code completion task should include a decision-making mechanism to stop generating incorrect code.
This feature, to the best of our knowledge, is novel in the context of dynamic inference and addresses a critical demand for improving the quality and effectiveness of code completions.

To fill this gap, we design a dynamic inference method for the code completion task, named Stop\&Exit Controller (\method).
It not only skips the unnecessary computation for generating each token (denoted as Exit) but also timely ``gives up'' (i.e., stopping the generation of current and subsequent tokens and outputting the sequence of already generated tokens) when realizing the current input is insufficient for a correct prediction (denoted as Stop).
For a clearer understanding of \method's functionality, we resent two illustrative examples in \Cref{fig:example}.
In the scenario on the left-hand side, the model recognizes that continuing the generation would lead to an incorrect completion and, therefore, decides to STOP generating any further tokens. 
In the second scenario, the model successfully generates the correct completion with just a few layers and promptly decides to EXIT early.
Adapting from the existing methodologies in dynamic inference~\cite{Schuster2022ConfidentAL}, \method utilizes a linear classifier to make informed decisions as the input code context passes through each layer of the LCM during the generation process.
By incorporating \method into the process, fewer layers are involved in generating tokens, resulting in reduced computational overhead and generating fewer unhelpful suggestions. 
Importantly, the classifier used in \method has only thousands of parameters, contributing negligible computational cost to the overall inference process.

To sufficiently evaluate the effectiveness of \method, we conduct a comprehensive evaluation with multiple settings, including two programming languages (i.e., Java and Python), two widely used LCMs (i.e., GPT-2 and CodeGen), and tens of threshold settings.
Firstly, we measure the accuracy of the action classifier in predicting proper actions. 
The action classifier can achieve high Precision for both Stop and Exit, respectively 0.812 and 0.944 at the 0.9 threshold, which is necessary for preserving the LCM’s performance.
Next, we investigate the computational efficiency of the \method-enhanced LCMs.
Surprisingly, on average of all the settings, \method can skip 1.7 layers out of 16 layers of the LCMs and speeds up 11.2\% during the completion generation with only a 1.1\% ROUGE-L reduction.
More importantly, the Acceptance Rate of the completions is well preserved or even improved since \method can prevent the unhelpful part in a piece of code completion from being generated.
Among six evaluated threshold settings, only one aggressive setting achieves a slightly reduced Acceptance Rate, from 46.7\% to 44.9\%, but speeds up the inference by 71\%.

To the best of our knowledge, this is the first study and adaption of dynamic inference methods for LCM-based code completion systems.
Our contribution can be summarized as follow:
\begin{itemize}[leftmargin=*]
    \item An experimental investigation on the inference capability of intermediate layers of LCMs, which reveals the potential wastes caused by unnecessary computation and the consequences of wrongly predicted tokens.
    \item A dynamic inference method, Stop\&Exit Controller (\method), designed for the code completion task, to not only skip the unnecessary computation but also prevent the LCMs from generating unhelpful completions.
    \item A comprehensive evaluation, demonstrating the feasibility and effectiveness of \method.
\end{itemize}

To facilitate future research and industrial practices, the code of \method is available at \url{https://github.com/v587su/SEC}.
\section{Background on Transformer-based Code completion}
\label{sec:transformer}

\begin{figure}[t]
\centerline{\includegraphics[width=0.95\columnwidth]{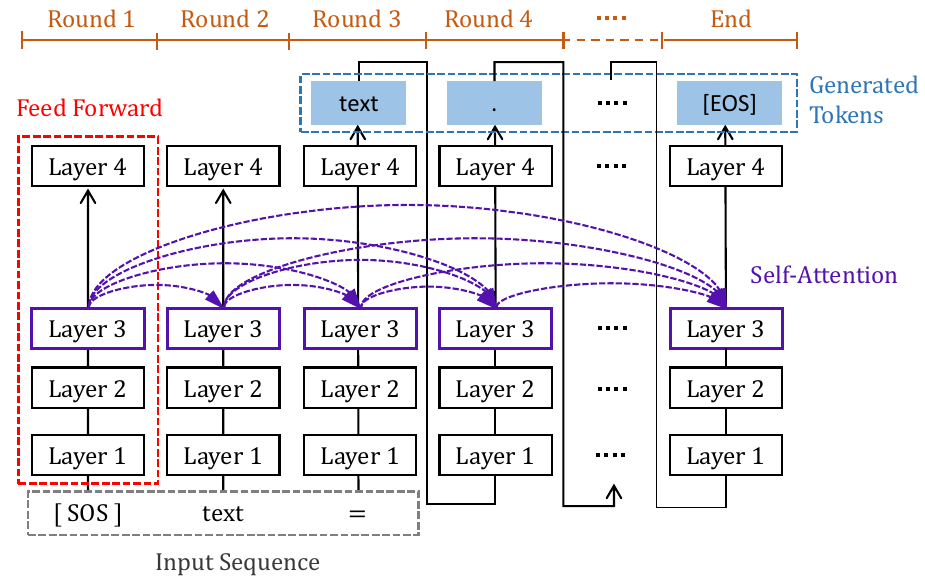}}
\caption{Demonstration of the typical generation process of a 4-layer Transformer, where the model generates four tokens for the input sequence using four steps. \emph{[SOS]} and \emph{[EOS]} are special tokens respectively indicating the start of the input sequence and the end of the generation.}
\label{fig:background}
\end{figure}

In this section, we provide a brief recap of the working mechanism of Transformer~\cite{Vaswani2017AttentionIA} in the context of code completion, given that Transformer and its variants have emerged as the dominant architecture for LCMs.
As most LCMs for code completion are decoder-only, we will focus on its decoder network and only reveal details relevant to our method, for the sake of simplicity.
Readers can refer to~\cite{Vaswani2017AttentionIA} for a comprehensive understanding of Transformer.

The decoder of a Transformer model consists of stacked layers, and each layer contains a self-attention component that maintains a set of attention states. \Cref{fig:background} provides an illustrative example of a Transformer with four stacked layers, showcasing its code completion process for the context ``\texttt{text =}''. Before inputting the context into the Transformer, it undergoes tokenization, forming a sequence of tokens. 
Additionally, a special token, \texttt{[SOS]}, is appended to mark the sequence's beginning.
The tokens are sequentially processed by the Transformer, triggering feed-forward propagation across all layers. Each token's representation is computed and output by the final layer (Layer 4 in the example). Once all tokens in the code context are processed, the representation of the last token is fed to the Language Model Head (LM Head), a softmax-normalized linear classifier, to predict the next token. The newly generated token undergoes the same feed-forward propagation, predicting one more token. This process continues until the token \texttt{[EOS]} is generated, indicating the end of the completion, or a pre-set maximum number of tokens is reached.

Now we explain the self-attention mechanism in detail. 
This mechanism facilitates the propagation of context information not only from shallow layers to deep layers but also across different rounds in a layer-specific manner.
In each layer, there are two types of states: the attention state and the hidden state. During each round, the layer calculates its attention state using the recorded attention states from all historical rounds of the same layer, along with the hidden state of its preceding layer (except for the first layer, which uses the token embedding instead). 
Subsequently, the layer computes its hidden state using the updated attention state and the hidden state of its preceding layer.
The final hidden state of the last layer serves as the representation of all tokens consumed up to that point. This representation is known as a contextualized representation, containing the embedded context information.
In~\Cref{fig:background}, we visualize how the attention states of Layer 3 propagate across different feed-forward rounds with dashed purple arrows.
Gaining a deep understanding of how the context information is exchanged and updated within the self-attention mechanism is important for designing dynamic inference methods, as they also rely on the attention state and the hidden state to effectively control the inference process during runtime.

\section{Empirical Investigation}
\label{sec:explore}

\begin{table}[t]\centering
\setlength\tabcolsep{2pt}
\caption{An investigation of the number of layers required to correctly predict the next token for a given code context.}
\begin{tabular}{|c|c|c|c|c|c|} 
\hline
\textbf{Layers} & Failed & 1 Layer & 2-5 Layers & 6-11 Layers & 12 Layer \\ 
\hline
\textbf{Proportion} & 14.5\% & 54.4\% & 9.0\% & 19.2\% & 0.9\% \\
\hline
\end{tabular}
\label{tab:investigate}\vspace{-4mm}
\end{table}

The fundamental assumption of dynamic inference lies in the notion that certain layers in the model are not essential for accurate predictions on some inputs. 
By dynamically allocating the necessary layers during the inference stage, we can significantly reduce the computational cost associated with redundant layers. 
The more redundant layers we can identify, the greater the potential cost savings, making dynamic inference particularly effective in such scenarios.
Despite the promising prospects of dynamic inference, its feasibility in the context of code completion tasks remains unexplored. To address this gap, we conducted an empirical study on a widely used LCM, GPT-2~\cite{Radford2019LanguageMA} (224M parameters, 12 layers), fine-tuned on a code dataset, COFIC~\cite{Sun2022OnTI}.
COFIC comprises 764,985 Java code snippets for training and 84,998 for conducting the investigation.
The readers can refer to ~\Cref{sec:setting} for more details about the model and dataset. 
Specifically, our study explores the following two Investigation Research Questions (IRQs): 

\smallskip \noindent \textbf{IRQ1:} For different inputs, how many layers of the LCM are indispensable to yield correct predictions?

\smallskip \noindent \textbf{IRQ2:} Is it advantageous to continue code completion after a wrong token has been generated?

IRQ2 is a natural follow-up to IRQ1, as it examines the quality of the generated code following a wrong token prediction. 
In other words, it assesses whether predicting a wrong token serves as an indicator of the subsequent code generation's quality. 
Through addressing these research questions, we aim to shed light on the potential benefits and limitations of dynamic inference for code completion tasks.

\subsection{IRQ 1: Number of Indispensable Layers}
\label{sec:inv-rq1}

To address IRQ1, we evaluate GPT-2 on a simplified task, next token prediction, a widely used task for pre-training LCMs~\cite{li2023starcoder,Zheng2023CodeGeeXAP, Nijkamp2022CodeGenAO}. 
In this task, our objective is to find the minimum number of layers in the GPT-2 model necessary to achieve accurate predictions for the next token in code completion.
To enable each layer to provide a prediction for the next token, we associate each intermediate layer of GPT-2 with an intermediate LM head.
The LM head takes in the hidden state of the corresponding layer and predicts the next token. We train the LM heads jointly using a summed loss, such that the prediction performance of every head is sufficiently optimized without favoring any specific layer.
Further details about the design and training of these LM heads are available in \Cref{sec:inter_lm_head}.
Intuitively, we hypothesize that the hidden state of a deeper layer provides more informative cues than a shallow one.
In other words, if a shallow layer correctly predicts the next token, it indicates sufficient information has been captured.
Therefore, during the prediction on the testing set, we report the shallowest layer that achieves the correct answer as the minimum number of layers indispensable for the inference.
In cases where all 12 layers produce incorrect results, we mark the prediction for that token as failed.

The results of this exploratory experiment are reported in \Cref{tab:investigate}.
Surprisingly, \emph{only} 0.9\% of tokens require the full computing capacity (i.e., 12 layers) of GPT-2, 
while 54.4\% of tokens can be correctly predicted using the first layer solely.
This finding highlights the inefficiency of the fixed ``best-effort'' inference mechanism of the LCM, leading to substantial computational waste.
The average number of required layers per token is 2.5 layers,
and consequently, around 80\% of computation can be saved if we could perfectly decide how many layers should be used to predict each token.
This represents the theoretical upper bound of computational resources that can be saved for GPT-2 while preserving its performance.
However, achieving perfect decisions on the minimum number of layers for every token in the diverse and complex code contexts is extremely challenging.
Despite the difficulty, even solutions with modest performance could lead to significant computational savings.
Moreover, the predictions for 14.5\% of tokens fail, revealing the high likeliness of the LCM to produce incorrect tokens.
Notably, it is the proportion for one single token, while a piece of code completion usually composes tens of tokens.
However, we are unclear whether such incorrect predictions are still helpful to developers, which will be investigated in IRQ2.

\begin{tcolorbox}[size=title]
{\textbf{Finding 1:}}
On average, only 2.5 layers, instead of 12 layers, are actually required by the LCM, where around 80\% of the computational resources are unnecessary. 
\end{tcolorbox}

\subsection{IRQ2: Helpfulness of the Completion after Generating a Wrong Token}
\label{sec:inv-rq2}
We further study the effects of the 14.5\% failed predictions in IRQ1 on the helpfulness of the code completions using the same dataset and model.
For each failed token observed in IRQ1, we additionally let the model finish the whole completion, forming a group of continued completions.
A control group is derived by completing randomly selected code contexts in the testing set.
By measuring and comparing the helpfulness of the completions in these two groups, we can understand the effects of these failed predictions.
According to the prior study \cite{Ziegler2022ProductivityAO}, among tens of alternative measures, the Acceptance Rate (the possibility of a piece of code completion being accepted by the developer) of code completion is the best metric for measuring the helpfulness of the completions in practice.
Thus, a manual inspection is conducted to measure the Acceptance Rate instead of computing automated metrics for accuracy. 

Specifically, two Ph.D. students with over 3 years of Java experience are recruited as volunteers for this inspection.
Given the code context, generated code completion, and the ground truth, they are asked to mark if the generated code completion can be accepted as a continuation of the code context.
The ground truth is provided to help annotators accurately understand the exact intent of the code context. Therefore, syntactically correct but unintended completions are also labeled as unaccepted.
The accepted completions should exactly match or achieve the same intention as the ground truth (the next 10 tokens in the origin code snippet).
The inspection checks two groups of completions: the continued completions after wrongly predicted tokens and the ones for random code context samples as a controlled baseline.
As it is non-trivial to manually annotate all the completions, we sample a subset (383 completions) from each group, where
the sample size of each group is statistically computed at a 95\% confidence level according to \cite{cochran1977sampling}.
During the annotation, the two annotators first independently annotate each subset and then discuss to reconcile all disagreements, thus mitigating the bias.
The proportion of accepted completions is computed as the Acceptance Rate.

The results of the annotation reveal that the Acceptance Rate of the continued completions after wrongly predicted tokens is only 4.2\%, which is significantly lower than that of the control group, 46.7\%.
An important conclusion can be drawn that if a specific token is wrongly generated, its subsequent generation is rarely helpful.
It also motivates us to reduce unhelpful completions by terminating the generation for tokens that are foreseen to be incorrect.
Though the helpfulness of the code completions cannot be comprehensively assessed solely based on the correctness of the first token,
it still provides a promising strategy for reducing unhelpful completions.
The evaluation of our proposed method (cf.~\Cref{sec:rq3}) demonstrates the effectiveness of this strategy.

\begin{tcolorbox}[size=title]
{\textbf{Finding 2:}}
The continued completions after wrongly predicted tokens are rarely helpful, with an Acceptance Rate of only 4.2\%. 
\end{tcolorbox}
\section{Proposed Method}

\begin{figure*}[t]
\centerline{\includegraphics[width=2\columnwidth]{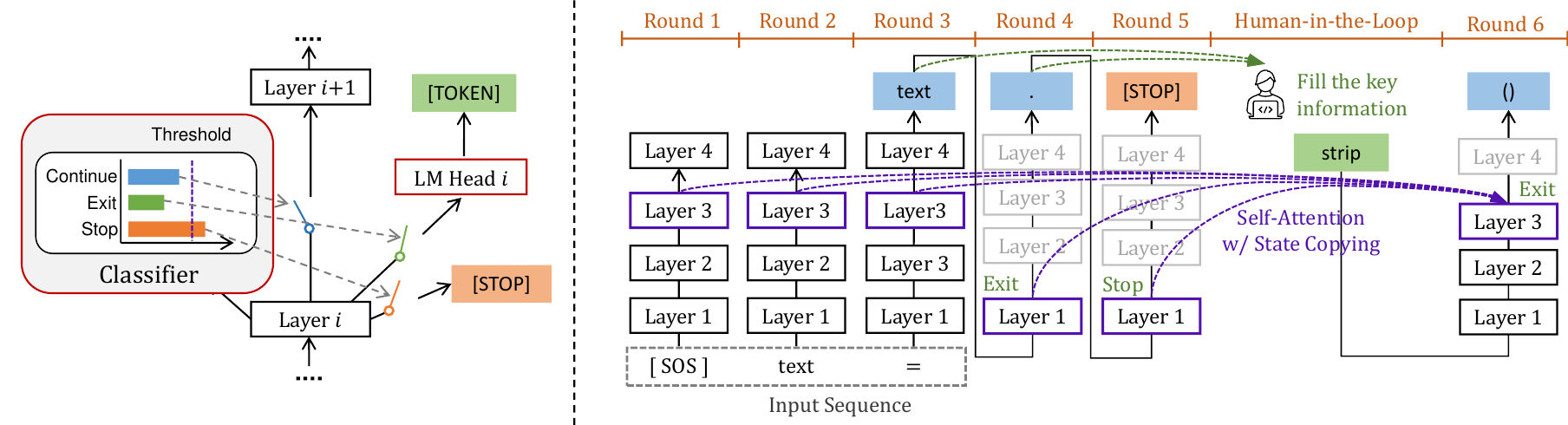}}
\caption{Demonstration of the working mechanism of \method. The left part shows how \method controls the inference using a classifier after Layer $i$ computing its hidden state. The right part showcases the generation process of a 4-layer \method-enhanced LCM, where \method exits at Round 4\&6 and stops at Round 5. The layers skipped by \method are in grey color.}
\label{fig:overview}
\end{figure*}

This section presents our dynamic inference framework, Stop\&Exit Controller (\method), to help the neural code completion model wisely allocate its computational resources during inference.
Below, we first describe the working mechanism of \method during each inference (\Cref{sec:sec}) and its training method (\Cref{sec:training}), then introduce the generation process of an \method-enhanced LCM (\Cref{sec:generation}).

\subsection{Stop\&Exit Controller for Dynamic Inference}
\label{sec:sec}

\method embeds a controller between the layers of the LCM to wisely decide when to skip the rest computation or timely ``give up'' (i.e., stopping the prediction of current and subsequent tokens).
As recapped in ~\Cref{sec:transformer}, in each Round, the LCM predicts one next token for the current input sequence.
Assume an input token sequence $x = (w_1,...,w_p,w_{p+1}...,w_t)$, where the first $p$ tokens form the input code context and the last $t-p$ tokens are newly generated so far.
The inference Round of an LCM without \method predicts the next token $w_{t+1}$ using all of its layers for the token sequence $x$.
Namely, $x$ goes through the $n$ layers $L_1,...,L_n$ of the LCM during which for each token $w_j \in x$, a hidden state $h_j^i$ is computed at the layer $L_i$, where $\textbf{h}_{j}^{0} = Embedding(w_j)$ and $ \textbf{h}_{j}^{i} = L_i(\textbf{h}_{j}^{i-1})\ \mbox{for} \ i>0$.
Next, an LM head $Head_n$ predicts the next token $w_{t+1}$ for the token sequence $x$ based on the hidden state $\textbf{h}_t^{n}$ of the last token $w_{t}$ at the last layer $L_n$, formally: $w_{t+1} = argmax(Head_n(\textbf{h}_{t}^{n}))$.
\method enhances this inference process by predicting the subsequent actions for some specified layers of the LCM to behave, using its action classifier $C$.
For brevity but without loss of generality, we assume \method is set to control the subsequent behavior of every intermediate layer of the LCM. 
We remark that \method maintains \emph{only} one classifier for all layers, instead of multiple action classifiers, to keep the \method-enhanced LCM as lightweight as possible.
As demonstrated on left of ~\Cref{fig:overview}, after the layer $L_i$ finishes its computation, the hidden state $\textbf{h}_{t}^{i}$ of the last token $w_t$ will be fed to the action classifier $C$ to decide the next action $A_i$ among three options: \emph{Stop} $A_s$, \emph{Exit} $A_e$, and \emph{Continue} $A_c$, computed as follows:
\begin{align*}
\textbf{p}_{i} &= softmax(C(\textbf{h}_{t}^{i})), \\
{A}_{i} &= 
\begin{cases}
 {A}_{s}, & \text{if } argmax(\textbf{p}_{i})={A}_{s} \text{ and } max(\textbf{p}_{i})\textgreater\alpha;\\
 {A}_{e}, & \text{if } argmax(\textbf{p}_{i})={A}_{e} \text{ and } max(\textbf{p}_{i})\textgreater\beta;\\
 {A}_{c}, & \text{otherwise};\\
\end{cases}
\end{align*}
where $\alpha$ and $\beta$ are two user-defined thresholds to respectively limit the minimal confidence score of $A_s$ and $A_e$. 
A higher threshold makes the decision of the action classifier $C$ more conservative.

The corresponding behavior of the inference process for each action predicted by \method is described as follows:

\smallskip
\noindent \textbf{Stop}: 
\emph{Stop} is to immediately stop the current round and subsequent generation rounds for the input code context.
Specifically, it interrupts the ongoing inference and returns a special token \textit{[STOP]}.
After receiving this special token, the generation process will be stopped and the sequence of tokens that have been generated so far will be displayed to the user.
\emph{Stop} is designed for the case when the rest yet-to-generate part of the completion is unhelpful.
Thus, subsequent tokens will not be generated to reduce the harm of unhelpful completions to development productivity and preserve computational resources.

\smallskip
\noindent \textbf{Exit}: 
\emph{Exit} is to generate the next token with the hidden state of the current layer and skip the rest layers of the current inference.
It is a widely investigated action for accelerating model inference~\cite{Zhou2020BERTLP}.
Predicting the next token with the hidden state relies on an LM head, but
the inherent LM head $Head_n$ of the LCM is \emph{only} applicable to the hidden state of the last layer. Thus, \method pairs an additional LM head $Head_i$ for each intermediate layer $L_i$ to fulfill the prediction.
We will describe their training in~\Cref{sec:inter_lm_head}.
Once \emph{Exit} is predicted by the classifier $C$ after the layer $L_i$, the current hidden state $\textbf{h}_{t}^{i}$ will be directly fed to the LM head $Head_{i}$ of this layer to compute the next token $w_{t+1}$, where
$w_{t+1} = argmax(Head_i(\textbf{h}_{t}^{i}))$.
Theoretically, if the classifier $C$ can properly make the decision of \emph{Exit}, the performance of the LCM will be well preserved.
Considering the computation for the rest $(n-i)$ layers is saved, the inference of the LCM can be faster and cheaper without sacrificing its performance.

\smallskip
\noindent \textbf{Continue}: 
\emph{Continue} is to proceed to compute the hidden state $\textbf{h}_{t}^{i+1}$ of the next layer $L_{i+1}$ by: $\textbf{h}_{t}^{i+1} = L_{i+1}(\textbf{h}_{t}^{i})$.
It is the default action for an LCM.
\method takes this action when neither \emph{Exit} nor \emph{Stop} is predicted by the classifier $C$.
After the new hidden state $\textbf{h}_{t}^{i+1}$ is computed, a new action may be predicted by the classifier $C$ until the inference ends at the final layer.

\begin{figure*}[t]
\centerline{\includegraphics[width=1.7\columnwidth]{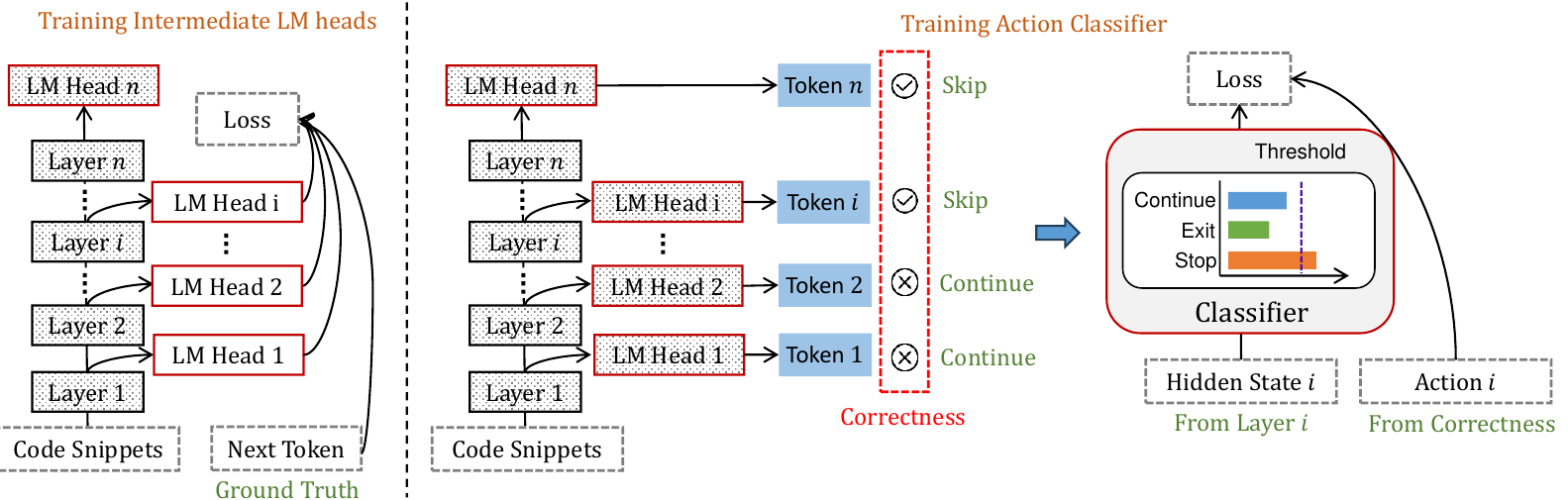}}
\caption{The training process required for integrating \method for an LCM (contains $n$ layers). The Layers and LM Heads with shadows indicate that they are well-trained and their parameters are frozen during training.}
\label{fig:training}
\end{figure*}

\subsection{Training \method}
\label{sec:training}
Among the components of \method, the action classifier $C$ and the intermediate LM heads 
$\{Head_1,...,Head_{n-1}\}$ are neural networks that need to be trained.
We first introduce the training of the intermediate LM heads, then the action classifier.

\subsubsection{Training Intermediate LM heads}
\label{sec:inter_lm_head}
To enable each layer to produce a prediction, each intermediate layer $L_i$ of the LCM is paired with an intermediate LM head $Head_i$ which predicts the next token based on the hidden state of the layer $L_i$.
Note that the inherent LM head $Head_n$ of the last layer $L_n$ is already trained during the training of the LCM, and thus does not need to be re-trained.
These intermediate LM heads are one-layer linear classifiers that are integrated into the well-trained LCM and independently trained without affecting the inherent parameters of the LCM.
Thus, compared with the training of the LCM, the cost of training such simple linear classifiers is negligible.
In detail, the training of these intermediate LM heads is similar to that of the LM head $Head_n$ except for involved layers.
Moreover, we use the same dataset, i.e., the corpus of source code files, for the training.
Thus, no additional training data or annotation efforts are required for training the intermediate LM heads.
The training process of intermediate LM heads is visualized on the left of~\Cref{fig:training}.
Given a tokenized code snippet $x = (w_1,...,w_p)$ from the training dataset, a training sample 
is derived for each token $w_j$ except for the last one $w_p$.
One training sample for $Head_i$ is a 
pair $({\bf h}_j^i, w_{j+1})$, where ${\bf h}_j^i$ is the hidden state of the layer $L_i$ computed by the LCM and $w_{j+1}$ is the next token of the token $w_j$ to serve as the ground truth.
With a training sample $({\bf h}_j^i, w_{j+1})$, each intermediate LM head computes a loss values $l$ between its predictions and the ground truth $w_{j+1}$,
where $l = Loss(Head_i(\textbf{h}_j^i)), w_{j+1})$.
The gradients are back-propagated to update the parameters of the intermediate LM head $Head_i$.

\subsubsection{Training An Action Classifier}
During the inference process of an \method-enhanced LCM, its action classifier $C$ predicts the subsequent action for each specified intermediate layer based on its hidden state.
Its supervised training requires a specific training dataset, the construction of which amounts to assigning a proper label, i.e., a subsequent action, to the hidden state of each specified intermediate layer.
Thus, the training sample is a pair $({\bf h}_j^i, A_j^i)$, where the hidden state ${\bf h}_j^i$ is obtained following the same way as that of intermediate LM heads introduced in~\Cref{sec:inter_lm_head} and the action $A_j^i$ is assigned according to the performance of all the layers.
As shown in the right of~\Cref{fig:training}, we first predict a next token $w_{j+1}^i$ with the hidden state ${\bf h}_j^i$ of each layer using the well-trained intermediate LM head $Head_i$ (LM Head $Head_n$ for the hidden state ${\bf h}_j^n$ of the last layer).
Then, we assign the label for each hidden state ${\bf h}_j^i$ according to the following rules: 1) \emph{Stop} when the layer $L_i$ and all its deeper layers cannot correctly predict the next token, 2) \emph{Exit} when the layer $L_i$ can predict the correct token, 3) \emph{Continue} when neither \emph{Stop} nor \emph{Exit} is applicable.
Formally, the assigned action $A_j^i$ for the hidden state $h_j^i$ is defined as:
\begin{equation}
    A_j^i = 
\begin{cases}
 {A}_{s}, & \text{if } w_{j+1}^k \neq w_{j+1}, \  \forall k \in \{i,..., n\};\\
 {A}_{e}, & \text{if   } w_{j+1}^i = w_{j+1};\\
 {A}_{c}, & \text{others}.
\end{cases}\nonumber
\end{equation}
Intuitively, the condition $w_{j+1}^k \neq w_{j+1}$ $\forall k \in \{i,..., n\}$ means that none of the rest layers can correctly predict the next token with their LM heads, thus the \emph{stop} action ${A}_{s}$ should be predicted by the action classifier $C$ to stop the current and subsequent inferences.
The condition $w_{j+1}^i = w_{j+1}$ means that
the hidden state ${\bf h}_j^i$ of the layer $L_i$ is sufficient for the LM head $Head_i$ to predict the correct next token, thus the rest layers should be saved in the current inference by assigning the \emph{Exit} action ${A}_{e}$.

With the sample pairs $({\bf h}_j^i, A_j^i)$ constructed as above, we train a one-layer linear classifier as the action classifier $C$ by minimizing a weighted average cross-entropy loss, which is computed by:
\begin{equation*}
    L = \sum^n_{i=1}m_i \cdot CELoss(C({\bf h}_j^i),A_j^i) \ \mbox{ subject to }\sum^n_{i=1}m_i = 1.
\end{equation*}
Following previous studies for dynamic inference~\cite{Schuster2022ConfidentAL, Zhou2020BERTLP}, we set $m_i = i/\sum^n_{k=1}k$ to favor deep layers, which helps preserve the model's performance during inference.

\subsection{The Generation Process of LCMs with \method}
\label{sec:generation}

An LCM generates a piece of code completion in a token-by-token manner, where one single inference process is executed to generate one token.
However, \method is integrated into the inference process without a global understanding of the generation of the whole completion.
Thus, some additional measures are desired to preserve the global performance of the LCM, which are introduced as follows:

\subsubsection{State Copying for \emph{Exit}}
\label{sec:copy}
As mentioned in \Cref{sec:transformer}, the computation of a hidden state requires the attention state of the same layer computed at the current round and the previous rounds.
These attention states are recorded to be reused in subsequent rounds.
However, if the inference for the token $w_j$ exits or stops at the layer $L_i$, the attention states and hidden states of the rest $(n-i)$ layers will not be computed, and thus are not
available when subsequent rounds require more layers.
For a better understanding of this issue, let us consider the example shown in ~\Cref{fig:overview} highlighted with purple.
Layer 3 at Round 6 requires the attention states computed at the same layer for all the previous rounds, 
but \emph{Exit} and \emph{Stop} respectively happen in Round 4 and Round 5 with no attention states computed at Layer 3, thus blocking the current inference.
To address this issue, we adopt a state copying mechanism.
Specifically, given the hidden state $h_j^i$ at the exited or stopped layer, we set all the hidden states $(h_j^{i+1},...,h_j^{n})$ of the rest layers as $h_j^i$ so that the attention states of these layers can be computed using the copied hidden state $h_j^i$, i.e., the hidden state of its preceding layer.
Thus, for Round 4 and Round 5 in the example, their attention states of Layer 3 were supposed to be computed based on the hidden states of Layer 2 (which are not computed due to \emph{Exit} and \emph{Stop}), but with the state copying mechanism, they are instead respectively computed with the hidden state of Layer 1 at Round 2 and Layer 1 at Round 3.
Though, intuitively, the state copying mechanism may introduce errors to the generation process,
our experiments in \Cref{sec:rq2} investigated such errors and found \method using state copying can achieve high efficiency with negligible sacrifice on accuracy.
Notably, we compute the attention states of deeper layers using the copied hidden state instead of directly copying the attention states of the exited layer since the attention states computed by different layers are not aligned.
Compared with the full computation of a layer, the computation of its attention state is cheap since its further steps, e.g., computing the hidden state, are avoided.

\subsubsection{Human-in-the-loop Code Completion by \emph{Stop}}
Recall that the generation process of an \method-enhanced LCM will be immediately interrupted when \emph{Stop} is predicted by \method,
namely, if the $i$-th generated token is [STOP], the tokens generated so far $ (w_1,...,w_{i-1})$ will be fed back to the developer, and no further tokens will be generated.
Thus, as demonstrated in~\Cref{fig:example}, \emph{Stop} can still retain the valuable part of the completion for the developer to reuse, providing a new form of interaction to developers.
Specifically, given the partial code completion, developers can choose to accept the partial completion and continue coding, dismiss the partial completion, or issue the completion request for the rest part anyway.
For example, in Round 5 of~\Cref{fig:overview}, it is hard for the LCM to decide which API to use without enough information in the context, and thus \method stops the generation to avoid the cost of producing a random guess.
Further, the developer fills in the key information, i.e., the API name \textit{strip}, enabling the generation to continue. 
We regard this as a new paradigm of how developers interact with the code completion system.
Traditional neural code completion systems provide automated line-level suggestions regardless of their helpfulness.
The generation process of these suggestions is out of the control of the developers, completely relying on the capabilities of the LCM itself.
In contrast, \emph{Stop} fosters a collaborative environment between developers and code completion systems, where human intelligence complements machine learning capabilities.
Developers can provide crucial information that the system may have difficulty inferring, leading to more accurate and contextually relevant completions.
Additionally, the ability to stop the generation process at critical points prevents the generation of unhelpful or incorrect suggestions, thus saving developers from having to review and dismiss irrelevant completions.
With \method as a pioneering example, we call for more investigation in this paradigm towards more interactive and human-in-the-loop code completion systems.
 \section{Experiment Setup}
\label{sec:setting}

In this section, we present the experimental setup of the datasets and models, and the evaluation metrics used throughout the experiments for \method,
to answer the following Experimental Research Questions (ERQ): 

\noindent \textbf{ERQ1:} How accurate is the action classifier of \method? 

\noindent \textbf{ERQ2:} How much computation can be saved by \method? 

\noindent \textbf{ERQ3:} How is the quality of the completions generated by \method-enhanced LCM?

\subsection{Large Code Models}
We adopt two widely used LCMs in our research community, GPT-2 and CodeGen, to fulfill the code completion tasks.

\smallskip \noindent \textbf{GPT-2}: GPT-2~\cite{Radford2019LanguageMA} is a pre-trained large language model 
and has been used by the commercial code completion system Tabnine~\cite{tabnine}.
We fine-tune this model with our code datasets for the code completion task. 

\smallskip \noindent \textbf{CodeGen}: CodeGen, proposed by Salesfore~\cite{Nijkamp2022CodeGenAO}, is an open-source competitive LCM with Github Copilot.
It has been trained on a code corpus consisting of multiple programming languages 
and is capable of the code completion task, thus can be used without further fine-tuning in our experiments.


\subsection{Datasets}
\label{sec:datasets}
In this work, we evaluate \method for Python and Java, though \method is generic and applicable to other programming languages.
Recall that \method is designed for the code completion task, where the inputs to the LCMs are usually unfinished code.
Therefore, we use code snippet datasets instead of the popular benchmarks for LCM that are based on natural language requirements, such as HumanEval~\cite{Chen2021EvaluatingLL}.
The datasets for Java and Python are respectively:

\smallskip \noindent \textbf{Java}:
We use COFIC~\cite{Sun2022OnTI} as our Java dataset.
It is collected by extracting functions from Java repositories on Github, containing 849,984 code snippets (namely, function definitions).
As usual, we split the dataset into two proportions: 90\% for training and 10\% for testing.

\smallskip \noindent \textbf{Python}:
The Python dataset is CodeSearchNet (CSN)~\cite{Husain2019CodeSearchNetCE}.
It has been pre-split where the train and test split respectively contain 412,178 and 22,176 code snippets and are collected from Github repositories.

We use the train-split to fine-tune the GPT-2 model and train the \method for each LCM.
The test-split is used to evaluate the accuracy of \method and the performance of \method-enhanced LCMs.

\begin{figure*}[t]
\centerline{\includegraphics[width=1.8\columnwidth]{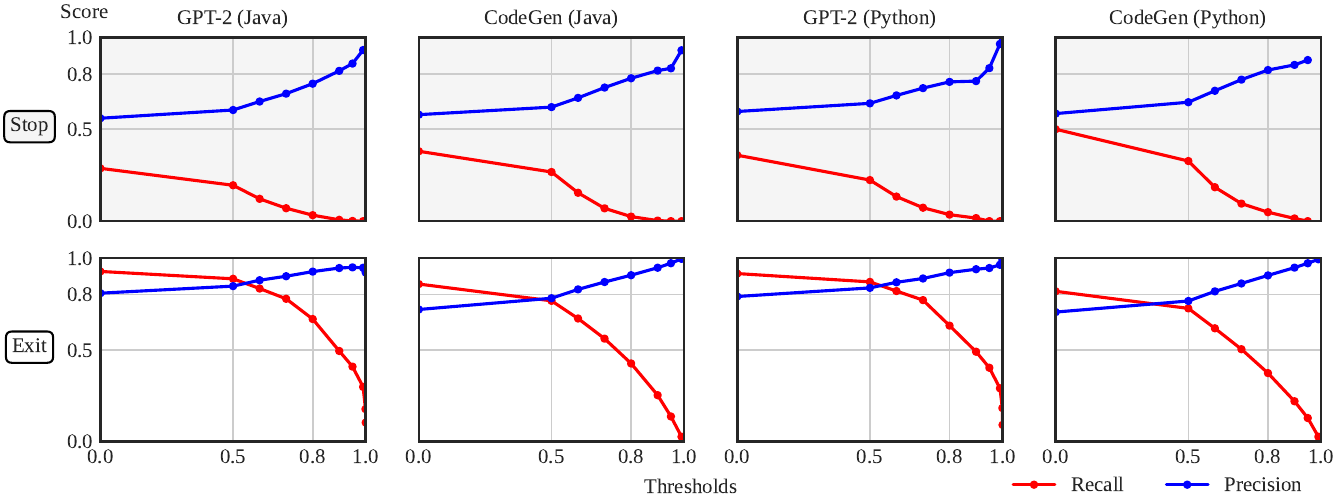}}
\caption{The accuracy of the action classifier of \method.}
\label{fig:rq1}
\end{figure*}

\subsection{Metrics}
\label{sec:metric}
Three widely used metrics are adopted in our evaluation.

\smallskip \noindent \textbf{Recall \& Precision:} 
    Recall \& Precision are two classical metrics for classification tasks, where Recall is the proportion of true samples that are predicted to be positive and Precision is the proportion of predicted samples that are correctly classified.
    We use them to measure the accuracy of  action classifiers.
    Since the cost of \method is negligible during the inference (an additional linear classier with thousands of parameters), it is enough to produce a positive impact in terms of efficiency or productivity by making a small number of right decisions.
    Thus, Precision is more important than Recall for the action classifier.

\smallskip \noindent \textbf{ROUGE-L:} ROUGE-L is a recall-oriented metric that looks for the longest common n-grams between the reference and the candidate. We use it to measure the performance of SEC-enhanced LCMs. 
We remark that Rouge-L is the best-performing one in an empirical study that compares 6 metrics for code generation models~\cite{Evtikhiev2022OutOT}.

\smallskip \noindent \textbf{Acceptance Rate:} Acceptance Rate indicates the proportion of completions that are marked as accepted, i.e., the number of accepted completions divided by the number of all completions.
According to \cite{Ziegler2022ProductivityAO}, among tens of alternative metrics, the Acceptance Rate of code completions is the best one for measuring the perceived developer productivity.

Notably, both ROUGE-L and Acceptance Rate are metrics to measure the quality of the completions but from different aspects.
ROUGE-L is automatically computed based on exactly matched tokens.
Such exact-match-based metrics may give improper credits to completions that dissatisfy the requirements of developers but are partially identical to the ground truth.
For example, the code completion that invokes an API in a wrong way can still receive a high ROUGE-L score, since the API name and some symbols may match the ground truth.
In contrast, Acceptance Rate relies on the judgment of human developers, which directly reflects the helpfulness of the completion in development.
However, Acceptance Rate is labor-intensive to be extensively computed.
Therefore, in the experiments, we use ROUGE-L to measure the sacrifice on completion quality for the efficiency improvement, where a large number of settings are evaluated, but judge the quality of completions generated by \method-enhanced LCMs with Acceptance Rate.

\subsection{Implementation Details}
GPT-2 is finetuned on a pre-trained version (124M parameters, 12 layers) with our training datasets and CodeGen is an off-the-shelf pre-trained model (350M parameters, 20 layers). 
Limited by our devices, we fail to deploy the larger versions of these LCMs.
We will analyze the impact of this limitation in~\Cref{sec:threat}.
\method is inserted after each intermediate layer in GPT-2,
but it is inserted after every three layers in CodeGen due to the limit of our computing resources.
We respectively set the maximum number of input tokens and generated tokens for one completion to 128 and 10, which is enough for LCMs to understand the code context and provide a full-line suggestion.
The finetuning of the GPT-2 model and the training of \method (including the action classifier and the intermediate LM heads) respectively take 10 and 5 epochs under a learning rate of 1e-5. 

\begin{table*}[t]\centering
\setlength\tabcolsep{4pt}
\caption{The results of the saving effects of \method-enhanced LCMs. The data in the gray background means that the tolerance size is too loose for this setting, where its results are also applicable under the tighter tolerance.}
\scalebox{0.85}{
\begin{tabular}{|c|c|c|c|c|c|c|c|c|c|cc|c|c|c|c|} 
\hline
\multirow{2}{*}{\textbf{Model}} & \multirow{2}{*}{\textbf{Lang.}} & \multirow{2}{*}{\textbf{Tolerance}} & \multicolumn{3}{c|}{\textbf{Exit only}} & \multicolumn{4}{c|}{\textbf{Stop only}} & \multicolumn{6}{c|}{\textbf{Both}} \\ 
\cline{4-16}
 &  &  & \textbf{\#Layers} & \textbf{Length} & \textbf{Speed} & \textbf{\#Layers} & \textbf{Length} & \textbf{Speed} & \textbf{\#Stops} & \multicolumn{2}{c|}{\textbf{ROUGE-L}} & \textbf{\#Layers} & \textbf{Length} & \textbf{Speed} & \textbf{\#Stops} \\ 
\hline
\multirow{10}{*}{GPT-2} & \multirow{5}{*}{Python} & Origin & 12.0~ & 9.7~ & $\times$1.00~ & 12.0~ & 9.7~ & $\times$1.00~ & 0.0\% & 0.598~ &  & 12.0~ & 9.7~ & $\times$1.00~ & 0.0\% \\
 &  & 1\% & 9.4~ & 9.7~ & \textbf{$\times$1.20~} & 12.0~ & 9.5~ & \textbf{$\times$1.01~} & 2.5\% & 0.591~ & 1.1\%$\downarrow$ & 9.4~ & 9.6~ & \textbf{$\times$1.18~} & 2.3\% \\
 &  & 5\% & 8.7~ & 9.7~ & \textbf{$\times$1.28~} & 11.8~ & 7.8~ & \textbf{$\times$1.20~} & 31.8\% & 0.562~ & 6.0\%$\downarrow$ & 8.7~ & 8.0~ & \textbf{$\times$1.48~} & 28.5\% \\
 &  & 10\% & 7.2~ & 9.8~ & \textbf{$\times$1.46~} & 11.6~ & 6.8~ & \textbf{$\times$1.34~} & 47.3\% & 0.516~ & 13.8\%$\downarrow$ & 7.4~ & 7.4~ & \textbf{$\times$1.84~} & 37.7\% \\
 &  & 20\% & 4.2~ & 9.9~ & \textbf{$\times$2.02~} & 10.9~ & 5.2~ & \textbf{$\times$1.67~} & 69.5\% & 0.421~ & 29.6\%$\downarrow$ & 4.7~ & 7.0~ & \textbf{$\times$2.97~} & 37.7\% \\ 
\cline{2-16}
 & \multirow{5}{*}{Java} & Origin & 12.0~ & 9.1~ & $\times$1.00~ & 12.0~ & 9.1~ & $\times$1.00~ & 0.0\% & 0.732~ &  & 12.0~ & 9.1~ & $\times$1.00~ & 0.0\% \\
 &  & 1\% & 11.1~ & 9.1~ & \textbf{$\times$1.06~} & 12.0~ & 8.6~ & \textbf{$\times$1.05~} & 8.7\% & 0.725~ & 0.9\%$\downarrow$ & 11.1~ & 8.6~ & \textbf{$\times$1.11~} & 8.7\% \\
 &  & 5\% & 9.1~ & 9.3~ & \textbf{$\times$1.22~} & 11.8~ & 7.6~ & \textbf{$\times$1.16~} & 24.9\% & 0.674~ & 7.9\%$\downarrow$ & 9.0~ & 7.8~ & \textbf{$\times$1.42~} & 24.9\% \\
 &  & 10\% & 8.3~ & 9.5~ & \textbf{$\times$1.29~} & 11.5~ & 6.4~ & \textbf{$\times$1.31~} & 42.9\% & 0.611~ & 16.5\%$\downarrow$ & 8.2~ & 6.8~ & \textbf{$\times$1.71~} & 44.8\% \\
 &  & 20\% & {\cellcolor[rgb]{0.871,0.871,0.871}}8.3~ & {\cellcolor[rgb]{0.871,0.871,0.871}}9.5~ & {\cellcolor[rgb]{0.871,0.871,0.871}}\textbf{$\times$1.29~} & {\cellcolor[rgb]{0.871,0.871,0.871}}11.5~ & {\cellcolor[rgb]{0.871,0.871,0.871}}6.4~ & {\cellcolor[rgb]{0.871,0.871,0.871}}\textbf{$\times$1.31~} & {\cellcolor[rgb]{0.871,0.871,0.871}}42.9\% & {\cellcolor[rgb]{0.871,0.871,0.871}}0.611~ & {\cellcolor[rgb]{0.871,0.871,0.871}}16.5\%$\downarrow$ & {\cellcolor[rgb]{0.871,0.871,0.871}}8.2~ & {\cellcolor[rgb]{0.871,0.871,0.871}}6.8~ & {\cellcolor[rgb]{0.871,0.871,0.871}}\textbf{$\times$1.71~} & {\cellcolor[rgb]{0.871,0.871,0.871}}44.8\% \\ 
\hline
\multirow{10}{*}{CodeGen} & \multirow{5}{*}{Python} & Origin & 20.0~ & 9.6~ & $\times$1.00~ & 20.0~ & 9.6~ & $\times$1.00~ & 0.0\% & 0.463~ &  & 20.0~ & 9.6~ & $\times$1.00~ & 0.0\% \\
 &  & 1\% & 18.5~ & 9.6~ & \textbf{$\times$1.03~} & 19.9~ & 8.8~ & \textbf{$\times$1.07~} & 12.8\% & 0.457~ & 1.3\%$\downarrow$ & 18.5~ & 8.8~ & \textbf{$\times$1.09~} & 12.8\% \\
 &  & 5\% & 17.4~ & 9.6~ & \textbf{$\times$1.05~} & 19.7~ & 7.8~ & \textbf{$\times$1.19~} & 29.7\% & 0.440~ & 4.9\%$\downarrow$ & 17.3~ & 7.7~ & \textbf{$\times$1.25~} & 30.5\% \\
 &  & 10\% & 15.4~ & 9.6~ & \textbf{$\times$1.09~} & 19.3~ & 6.5~ & \textbf{$\times$1.39~} & 49.6\% & 0.407~ & 12.1\%$\downarrow$ & 15.3~ & 6.4~ & \textbf{$\times$1.54~} & 52.0\% \\
 &  & 20\% & 12.5~ & 9.7~ & \textbf{$\times$1.15~} & 18.4~ & 5.1~ & \textbf{$\times$1.73~} & 68.1\% & 0.338~ & 26.9\%$\downarrow$ & 12.5~ & 5.0~ & \textbf{$\times$2.04~} & 73.1\% \\ 
\cline{2-16}
 & \multirow{5}{*}{Java} & Origin & 20.0~ & 8.6~ & $\times$1.00~ & 20.0~ & 8.6~ & $\times$1.00~ & 0.0\% & 0.623~ &  & 20.0~ & 8.6~ & $\times$1.00~ & 0.0\% \\
 &  & 1\% & 18.3~ & 8.6~ & \textbf{$\times$1.03~} & 20.0~ & 8.3~ & \textbf{$\times$1.04~} & 6.3\% & 0.615~ & 1.1\%$\downarrow$ & 18.3~ & 8.3~ & \textbf{$\times$1.07~} & 6.3\% \\
 &  & 5\% & 16.8~ & 8.6~ & \textbf{$\times$1.06~} & 19.7~ & 7.2~ & \textbf{$\times$1.17~} & 23.7\% & 0.588~ & 5.6\%$\downarrow$ & 16.7~ & 7.2~ & \textbf{$\times$1.24~} & 23.7\% \\
 &  & 10\% & 14.5~ & 8.7~ & \textbf{$\times$1.11~} & 19.3~ & 6.4~ & \textbf{$\times$1.29~} & 35.8\% & 0.537~ & 13.7\%$\downarrow$ & 14.3~ & 6.4~ & \textbf{$\times$1.44~} & 36.4\% \\
 &  & 20\% & 12.5~ & 8.8~ & \textbf{$\times$1.15~} & 18.4~ & 5.5~ & \textbf{$\times$1.47~} & 48.2\% & 0.476~ & 23.6\%$\downarrow$ & 12.2~ & 5.6~ & \textbf{$\times$1.72~} & 50.4\% \\
\hline
\end{tabular}}
\label{tab:ablation}
\end{table*}

\section{Experimental Results}
This section reports the experimental results and answers the research questions.
\subsection{ERQ1: Accuracy of \method}
\label{sec:rq1}

To answer ERQ1, we evaluate the accuracy of the action classifier, which is the key to the effectiveness of \method.
The accuracy is evaluated with four experimental settings in total, including two code datasets (i.e., Python and Java) and two LCMs (i.e., GPT-2 and CodeGen).
We respectively train and integrate the \method (both the classifier and the intermediate LM heads) into the LCM in each setting.
Each \method-enhanced model is used to predict the next token for each token of the code snippet from the test set, where the classifier's decisions at each layer are recorded.
It is noteworthy that the predicted actions are not executed in ERQ1 to observe the predictions from deeper layers.
Since \emph{Continue} is the default action of LCMs, we focus on the accuracy for \emph{Exit} and \emph{Stop} actions.
In each setting, we respectively run the evaluation under a wide range of thresholds $(0, 0.5, 0.6, 0.7, 0.8, 0.9, 0.95, 0.99, 0.999)$ for each action, covering from the most conservative setting to the most aggressive one with reasonable intervals.
The accuracy metrics (Recall and Precision) are recorded in the evaluation.

The results of the action classifier on the testing set under each setting are reported in~\Cref{fig:rq1}.
First, by setting a proper threshold, both \emph{Stop} and \emph{Exit} can achieve significantly high Precision, namely, the action classifier rarely makes incorrect predictions.
It is important for \method since, being cost-friendly, \method can produce a positive impact in terms of efficiency or productivity by conservatively making a small number of right decisions.
For example, the Precision of \emph{Stop} and \emph{Exit} at a threshold of 0.9 on average of all the settings are respectively 0.812 and 0.944.
Second, \emph{Exit} can still preserve a promising Recall at a threshold of 0.95, where the average Precision and Recall are respectively 0.959 and 0.269.
Interestingly, the Recall of \emph{Stop} is generally a little bit low.
In~\Cref{tab:ablation}, we also observed that the average number of layers used by \emph{Stop} is significantly higher than those used by \emph{Exit}, approaching the total number of layers in the LCM.
This indicates that stopping the inference immediately when one incorrect token is about to yield is quite a challenging task.
Computation with more layers is required for the classifier to make a precise decision.
We leave an in-depth analysis and improvement of this issue as an interesting future work.
Finally, as the threshold increases, for both actions under all the settings, Precision goes higher while Recall grows lower.
There exists a trade-off in setting the threshold, i.e., a more strict threshold brings higher precision while leaving more cases unaddressed.
Thus, it is recommended to train the action classifier and experimentally 
evaluate its performance with different thresholds to decide a proper one for the production environment.

\begin{tcolorbox}[size=title]
{\textbf{Answer to ERQ1:}}
The action classifier of \method can be set to achieve high Precision for \emph{Exit} while maintaining a promising level of Recall. While making a precise decision for the \emph{Stop} action is a bit more challenging, leaving space for further improving the recall metric.
\end{tcolorbox}

\subsection{ERQ2: Computational Efficiency}
\label{sec:rq2}

To answer ERQ2, the four \method-enhanced LCMs trained in~\Cref{sec:rq1} and their original LCMs are evaluated to compare their computational efficiency.
Since both \emph{Stop} and \emph{Exit} can affect the performance of the LCMs, we first conduct an ablation study to observe their impact on efficiency independently.
For each action, we respectively set its threshold with the same threshold range as the one in~\Cref{sec:rq1} and disable the other action for ablation.
In each setting, the \method-enhanced LCM (or original LCM) generates a piece of code completion for each given code context, where each code context is made by randomly splitting each code snippet in the test set into two parts: the first part is the code context being fed to the LCM, and the second part serves as the ground truth completion of the code context.
We observe the efficiency of these generation processes and compute multiple metrics, including \#Layers (the average number of layers involved in the computation for generating each token of each completion), Length (the average length of each generated completion), and Speed (the ratio of the average running speed for generating each completion compared with the original LCM).
ROUGE-L is also computed to observe the sacrifice for efficiency.
For \emph{Stop}, we additionally record \#Stops (the proportion of the completions where a \emph{Stop} happens).
By comparing the performance of \method-enhanced LCMs and their original LCMs, we can observe the effects of each action on efficiency.

For simplicity, we report the results by considering four tolerances on the reduction of ROUGE-L, i.e., $1\%, 5\%, 10\%, 20\%$, which correspond to four user preferences, but report the full results on our website~\cite{website}.
To be specific, among the results of all the thresholds, for each tolerance, we report the fastest result whose reduction on ROUGE-L falls within the range of the tolerance.
The results are shown in the left part of ~\Cref{tab:ablation}.
\emph{Exit}, designed to skip unnecessary computation during model inference, can significantly reduce the computation during the generation with a slight sacrifice on ROUGE-L.
When the tolerance is within 1\%, \emph{Exit} can skip 1.6 out of 16 layers ($\frac{12+20}{2}=16$ layers) and speed up the generation by 8.1\% on average of the four settings. 
When a 10\% sacrifice on ROUGE-L is acceptable, the speed of \method-enhanced LCM can be improved by 23.6\% compared with the original one. 
The saving effect of \emph{Stop} at low tolerance ($<1\%$) is also satisfying, achieving an average 4.1\% of speedup by stopping in 7.6\% of completions.
When the tolerance reaches 10\%, \emph{Stop} can achieve superior performance to \emph{Exit}, a 33.3\% speedup. 
Thus, both two actions are effective in boosting computational efficiency with acceptable sacrifices on ROUGE-L.

Further, we investigate the saving effects of \method by enabling both \emph{Stop} and \emph{Exit} at the same time.
We still adopt the thresholds of \emph{Stop} and \emph{Exit} for these tolerances in the ablation study, where the thresholds for the same tolerance are set together.
As shown in the right part of ~\Cref{tab:ablation}, the efficiency is significantly boosted with both actions enabled.
When using the thresholds for 1\% tolerance in the ablation study, \method skips 1.7 layers and increases the running speed by 11.2\%, while only reducing 1.1\% ROUGE-L on average of all settings.
It means that the same servers can additionally serve 11.2\% of user requests with only a 1.1\% reduction in ROUGE-L, which brings enormous economic value.
More sacrifices on ROUGE-L can bring better acceleration.
On average, the thresholds for 5\% and 10\% tolerance, respectively, reduce 6.1\% and 14.0\% ROUGE-L while increasing the speed by 34.9\% and 63.5\%.
Notably, with a 29.6\% ROUGE-L reduction, the speed of \method-enhanced GPT-2 (Java) is improved to almost three times its original version. 

\begin{tcolorbox}[size=title]
{\textbf{Answer to ERQ2:}}
\method can effectively conserve computational resources without significantly compromising the model's overall performance. It skips 10.6\% of computation with only a 1.1\% ROUGE-L reduction on average of all settings.
\end{tcolorbox}

\subsection{ERQ3: Quality of the Completions}
\label{sec:rq3}

\begin{table}[t]\centering
\setlength\tabcolsep{4pt}
\caption{The productivity measurement of \method-enhanced LCMs on Java dataset when both actions are enabled, AR=Acceptance Rate.}
\scalebox{0.86}{
\begin{tabular}{|c|c|cc|c|c|c|} 
\hline
\textbf{Model} & \textbf{Thresholds} & \multicolumn{2}{c|}{\textbf{ROUGE-L}} & \textbf{Length} & \textbf{\#Stops} & \textbf{AR} \\ 
\hline
\multirow{4}{*}{GPT-2} & Origin & 0.732~ &  & 9.7~ & 0.0\% & 46.7\% \\
 & 1\% Tolerance & 0.725~ & 0.9\%$\downarrow$ & 8.6~ & 11.1\% & \textbf{47.0\%} \\
 & 5\% Tolerance & 0.674~ & 7.9\%$\downarrow$ & 7.8~ & 31.8\% & \textbf{47.2\%} \\
 & 10\% Tolerance & 0.611~ & 16.5\%$\downarrow$ & 6.8~ & 57.3\% & 44.9\% \\ 
\hline
\multirow{4}{*}{CodeGen} & Origin & 0.623~ &  & 8.6~ & 0.0\% & 39.8\% \\
 & 1\% Tolerance & 0.615~ & 1.1\%$\downarrow$ & 8.3~ & 6.3\% & \textbf{40.1\%} \\
 & 5\% Tolerance & 0.588~ & 5.6\%$\downarrow$ & 7.2~ & 23.7\% & \textbf{44.1\% }\\
 & 10\% Tolerance & 0.537~ & 13.7\%$\downarrow$ & 6.4~ & 36.4\% & \textbf{46.2\%} \\
\hline
\end{tabular}
}
\label{tab:rq3}
\end{table}

To answer ERQ3, we investigate the quality of the completions generated by \method-enhanced LCM with both actions enabled to profile their helpfulness in practice.
We analyze the completions generated by the \method-enhanced LCMs and their original LCMs in ERQ2 for Java with the thresholds of three tolerances 1\%, 5\%, and 10\%.
As mentioned in~\Cref{sec:metric}, ROUGE-L may not be precise enough to measure the quality of completions in practice.
Thus, we conduct a manual inspection to measure the Acceptance Rate of the completions.
The inspection follows the same process described in~\Cref{sec:inv-rq2}.
There are eight groups of completions for the annotators to inspect, including six groups of completions respectively generated by \method-enhanced GPT-2 and CodeGen under the three thresholds and two groups
of completions respectively generated by the original GPT-2 and CodeGen.
By comparing the metrics between \method-enhanced LCMs and original LCMs, we can observe the effects of \method on the quality of the completions.

The results are summarized in~\Cref{tab:rq3}.
Under the conservative thresholds, the quality of the completions, measured by both ROUGE-L and Acceptance Rate, can be well preserved or improved.
For example, using the thresholds of 1\% Tolerance, the Acceptance Rate of both GPT-2 and CodeGen are respectively 47.0\% and 40.1\%, even higher than their original LCMs.
Considering the satisfying saving effects under the same thresholds (shown in ~\Cref{tab:ablation}), \method can be set to preserve the quality of the completions and accelerate the generation process at the same time.
To save more computations, the thresholds can be more aggressive, i.e., with higher Tolerance.
When the Tolerance is 10\%, on average of the two models, the number of tokens in a piece of completion drops 28.1\% and the ROUGE-L is reduced by 15.1\%.
However, there is no significant decrease or even a remarkable improvement in the acceptance rate, where GPT-2 drops from 46.7\% to 44.9\% and CodeGen increases from 39.8\% to 46.2\%.
Intuitively, the reduced ROUGE metric means that fewer tokens are generated correctly, which leads to a reluctance on the part of the user to accept the code completion.
However, the annotators found that \emph{Stop} can let many completions, that would have been discarded by the user if completely generated, be accepted.
Limited by the pages, we demonstrate several cases during the annotation on our website~\cite{website}, where after preventing the erroneous part by \emph{Stop}, the retained partial completion can still be helpful to the developers, thus being accepted. 
Therefore, though under aggressive thresholds, the completions generated by \method-enhanced LCMs can still be accepted with a comparable possibility.
In summary, we can adjust the thresholds used by SEC to strike a balance between completion quality and computational efficiency based on their specific requirements.

\begin{tcolorbox}[size=title]
{\textbf{Answer to ERQ3:}}
For LCM-based code completion systems that prioritize quality, \method can be set conservatively to effectively preserve the completion quality while significantly improving computational efficiency.
For those that prioritize efficiency, the completions generated by \method-enhanced LCMs can still maintain high acceptance rates, even with an acceptable trade-off in accuracy.
\end{tcolorbox}

\section{Related Work}
\smallskip
\noindent \textbf{Dynamic Inference for Large Models}
Instead of always using all the computations during inference, dynamic inference adaptively allocates computations for each input.
Multiple dynamic inference techniques~\cite{Xin2020DeeBERTDE, Hou2020DynaBERTDB,Xin2021BERxiTEE,Zhou2020BERTLP, Li2021CascadeBERTAI,Schuster2022ConfidentAL} have been proposed.
For example, Li et al.~\cite{Li2021CascadeBERTAI} accelerate the inference for classification tasks by invoking a series of models in a cascade manner. 
Schuster et al.~\cite{Schuster2022ConfidentAL} propose a framework for accelerating Transformer-based large language models.
However, to the best of our knowledge, dynamic inference has not been investigated for the code completion task, and the performance of dynamic inference on LCMs remains open. 
Furthermore, the unique nature of the code completion practices requires LCMs to prevent unhelpful completions, which is not covered in prior works.
\method is proposed to fill these gaps.

\smallskip
\noindent \textbf{Neural Code Completion}
The neural code completion task predicts the next few tokens for a given code context.
LCMs have emerged as a dominant player in this field due to their remarkable performance in generating accurate code completions.
Recently, a new LCM called AlphaCode~\cite{Li2022CompetitionlevelCG} has been released, garnering significant attention for its exceptional performance in programming competitions. 
Impressively, AlphaCode achieved a ranking in the top 54.3\% among over 5,000 human participants,
This success highlights the effectiveness of LCMs in code completion and their potential to revolutionize the field.
One noticeable trend in the development of LCMs is the constant increase in model size.
New LCMs such as StarCoder~\cite{li2023starcoder} and Pangu-Coder2~\cite{shen2023pangucoder2} have reached an impressive scale of 16 billion parameters, vastly surpassing earlier models like CodeBert~\cite{Feng2020CodeBERTAP} (125 million parameters) and CodeT5~\cite{Wang2021CodeT5IU} (770 million parameters). 
While larger models often exhibit superior performance, they also present a significant challenge in terms of inference workload, as more parameters bring higher computational requirements.
As the size of LCMs continues to grow, the inference costs associated with their deployment become a critical concern~\cite{chatgptCost, chien2023reducing}.
In this paper, we explore and adapt dynamic inference as a promising solution for cost-saving in code completion, paving the way for more resource-efficient and productive code completion applications.
\section{Threats to Validity}
\label{sec:threat}
\smallskip
\noindent\textbf{Limited experiments}.
Limited by our devices, we only conduct experiments on two LCMs: GPT-2 and CodeGen, both containing hundreds of millions of parameters.
Nevertheless, their architectures serve as prevalent templates for current LCMs, including newer models like StarCoder~\cite{li2023starcoder}, which achieved state-of-the-art performance by directly adopting the GPT-2 architecture with increased parameters and training samples.
Remarkably, while the model size is expanding, the fundamental issue of computation waste and unhelpful completions caused by the fixed inference process remains unaddressed. 
Recent research~\cite{Schaeffer2023AreEA} has even shown that the folklore emergent abilities of large models can be attributed to non-linear evaluation metrics.
Therefore, despite these limitations, the insights gained from our experiments remain valuable for understanding inefficiency concerns in other LCMs.

\smallskip
\noindent\textbf{Threshold}.
Our experiment does not recommend a universal ``silver bullet'' threshold that can satisfy the requirements of code completion systems in general.
An appropriate threshold should be determined based on the trade-off between cost-saving and accuracy degradation.
Therefore, by tuning the thresholds of \method, the systems can adjust the accuracy-efficiency trade-off to fit different devices and resource constraints without retraining the LCM, which provides real-world applications with more flexibility and adaptability.
Moreover, our method is independent of the LCM-based code completion systems and 
other cost-friendly endeavors, e.g., model compression~\cite{Shi2022CompressingPM, shi2023towards}, allowing system providers to experimentally evaluate the settings of thresholds based on their data logs accumulated in the production environment.

\smallskip
\noindent\textbf{Bias in annotation}
The Acceptance Rate, which measures the helpfulness of code completions, is derived from the evaluations conducted by two experienced human annotators.
Despite their expertise, human annotators may introduce inherent biases in their judgment, which could affect the accuracy of the acceptance rate.
To mitigate this threat, the annotators are provided with detailed instructions and discuss to resolve any disagreements and reach a consensus, further enhancing the reliability of their judgments. 
Besides, though being sampled with a statistically decided sample size, the small-scale groups for measuring the acceptance rate of the completions generated by \method-enhanced LCMs may introduce bias to the evaluation of our approach.

\section{Conclusion}
We first experimentally found that the LCMs actually need only a few layers in practice and the completions whose first token is incorrect are seldom helpful. 
Based on these findings, we have designed, to the best of our knowledge, the first dynamic inference framework for code completion models, namely \method, to optimize LCMs in terms of their computational costs and helpfulness.
The comprehensive evaluation of \method shows that it can significantly speed up the inference with negligible sacrifice on ROUGE-L.

\begin{acks}
This work is partially supported by the National Natural Science Foundation of China (NSFC) under Grant No. 62072309,
CAS Project for Young Scientists in Basic Research (YSBR-040),  ISCAS New Cultivation Project (ISCAS-PYFX-202201), 
ISCAS Fundamental Research Project (ISCAS-JCZD-202302), and
Open Foundation of Yunnan Key Laboratory of Software Engineering under Grant No.2023SE102.
\end{acks}

\balance
\bibliographystyle{ACM-Reference-Format}
\bibliography{sample-base}

\end{document}